\documentclass[10pt, conference, compsocconf]{IEEEtran}

\usepackage{tabu}

\ifCLASSINFOpdf
   \usepackage[pdftex]{graphicx}
\else
\fi

  \usepackage{listings}

\begin{document}
\newpage
\onecolumn
\begin{table*}
		\resizebox{0.9\linewidth}{!}{%
	\begin{tabular}{l}		
\textbf{Copyright © 2021 IEEE}	\\
\textbf{© 2021 IEEE. Personal use of this material is permitted. Permission from}\\
\textbf{IEEE must be obtained for all other uses, in any current or future media,}\\
\textbf{including reprinting/republishing this material for advertising or promotional}	\\
\textbf{purposes, creating new collective works, for resale or redistribution to servers}\\
\textbf{or lists, or reuse of any copyrighted component of this work in other works.}\\
\textbf{}\\  	[1ex] 	           
	\end{tabular}}
    
\end{table*}

\newpage
\twocolumn

\title{Container Orchestration on HPC Systems}
\author{\IEEEauthorblockN{Naweiluo Zhou$^1$, Yiannis Georgiou$^2$, Li Zhong$^1$, Huan Zhou$^1$, Marcin Pospieszny$^3$}
	\IEEEauthorblockA{High Performance Computing Center Stuttgart (HLRS), Germany$^1$\\
	Ryax Technologies, Lyon, France$^2$\\
	Institute of Bioorganic Chemistry of the Polish Academy of Sciences, \\
	Poznan Supercomputing and Networking Center,Poznan, Poland$^3$\\
	Email: naweiluo.zhou@hlrs.de, yiannis.georgiou@ryax.org\\ li.zhong@hlrs.de, huan.zhou@hlrs.de, marcin.pospieszny@man.poznan.pl
}
}

%\author{\IEEEauthorblockN{Authors Name/s per 1st Affiliation (Author)}
%	\IEEEauthorblockA{line 1 (of Affiliation): dept. name of organization\\
%		line 2: name of organization, acronyms acceptable\\
%		line 3: City, Country\\
%		line 4: Email: name@xyz.com}
%	\and
%	\IEEEauthorblockN{Authors Name/s per 2nd Affiliation (Author)}
%	\IEEEauthorblockA{line 1 (of Affiliation): dept. name of organization\\
%		line 2: name of organization, acronyms acceptable\\
%		line 3: City, Country\\
%		line 4: Email: name@xyz.com}
%}


\markboth{Journal of \LaTeX\ Class Files,~Vol.~11, No.~4, December~2012}%
{Shell \MakeLowercase{\textit{et al.}}: Bare Demo of IEEEtran.cls for Journals}



\maketitle

\begin{abstract}
Containerisation demonstrates its efficiency in application deployment in cloud computing. Containers can encapsulate complex programs with their dependencies in isolated environments, hence are being adopted in HPC clusters. HPC workload managers lack micro-services support and deeply integrated container management, as opposed to container orchestrators (\textit{e.g. Kubernetes}). We introduce Torque-Operator (a plugin) which serves as a bridge between HPC workload managers and container Orchestrators.

\end{abstract}	


% Singularity;
% \fi
\IEEEpeerreviewmaketitle

\section{Introduction}

\IEEEPARstart{C}loud computing demands high-portability. Containerisation ensures compatibility of applications and their environment by encapsulating applications with their libraries and configuration files \cite{8360359}, thus enables users to move and deploy programs easily among clusters. Containerisation is a virtualisation technology \cite{DBLP:journals/spe/RodriguezB19}. Rather than starting a holistically simulated OS on top of the host kernel as in a Virtual Machine (VM), a container only shares the host kernel. This feature makes containers more lightweight than VM. Containers are dedicated to run micro-services and one container mostly hosts one application. Nevertheless, containerised applications can become complex, \textit{e.g.} thousands of separate containers may be required in production. Production can benefit from container orchestrators that can provide efficient environment provisioning and auto-scaling.

High Performance Computing (HPC) systems are traditionally applied to perform large-scale financial and engineering simulation, which demands low-latency and high-throughput. The typical HPC jobs are large workloads that are often host-specific and hardware-specific. HPC systems are typically equipped with workload managers. A \textit{workload manager} is composed of a \textit{resource manager} and a \textit{job scheduler}. A resource manager \cite{Hovestadt2003} allocates resources (\textit{e.g.} CPU, memory), schedules jobs and guarantees no interference from other user processes. A job scheduler determines the job priorities, enforces resource limits and dispatch jobs to available nodes \cite{Klusacek2015}. Two main-stream workload managers are TORQUE \cite{Staples2006} and Slurm \cite{Jette02slurm:simple}. Slurm includes both resource managers and job schedulers, while originally Torque only incorporates resource managers and later extends with job schedulers. Overall, HPC workload managers lack micro-service supports and deeply-integrated container management capabilities in which container orchestrators manifest their efficiency.

We herein describe a plugin named \textit{Torque-Operator}. It serves as a bridge between the HPC workload manager \textit{Torque} and the container orchestrator \textit{Kubernetes} \cite{8457916}. Kubernetes has been widely adopted, as it has a rapidly growing community and ecosystem with plenty of platforms being developed upon it. Furthermore, we propose a testbed architecture composed of an HPC cluster and a big data cluster where Torque-Operator enables scheduling container jobs from the big data cluster to the HPC cluster. The rest of the paper is organised as follows. Firstly, Section~\ref{sec:related_work} briefly views the related work. Next, we describe the proposed architecture of our testbed and Torque-Operator in Section~\ref{sec:architecture}. Followed, some preliminary results are given in Section~\ref{sec:resulst}. Lastly, Section~\ref{sec:future_work} concludes this paper and proposes future work.

\section{Related Work}\label{sec:related_work}
Torque-Operator extends WLM-Operator \cite{wlmoperator2019} with Torque support. Both operators share similar mechanisms, \textit{i.e.} schedule container jobs from cloud clusters to HPC clusters, nevertheless, their implementation varies significantly as Torque and Slurm have different structures and parameters.   

WLM-Operator only allows submission of Slurm batch jobs wrapped in a Kubernetes \textit{yaml} file from a cluster managed by Kubernetes. It invokes Slurm binaries \textit{i.e.} \textit{sbatch}, \textit{scancel}, \textit{sacct} and \textit{scontol} to transfer and manage Slurm jobs to a Slurm cluster. The operator creates \textit{virtual nodes} which correspond to each Slurm partition, \textit{e.g.} one virtual node corresponds to one Slurm partition and contains the information of its corresponding partition. Virtual node is a concept in Kubernetes. It is not a real worker node, however, it enables users to connects Kubernetes to other APIs and allows developers to deploy \textit{pods} (a Kubernetes term) and containers with their own APIs. Jobs on the virtual node can be scheduled to the worker nodes. WLM-Operator creates a \textit{dummy pod} on the virtual node in order to transfer the Slurm batch job to a specific Slurm partition. When the batch job completes, another dummy pod is generated to transfer the results to the directory specified in the submitted yaml file.

In Kubernetes terminology, WLM-Operator creates a new \textit{object kind} \textit{i.e. Slurmjob}. The operator includes a service program \textit{red-box} that builds a gRPC proxy between Slurm and Kubernetes. gRPC proxy defines a service and implements a server and clients. The service defines the methods and their message types of responds and requests in a \textit{.proto} format file. The server implements: 1) the interfaces 2) and runs a gRPC server which listens to the requests from clients and dispatches them to the right services. The client defines the identical methods as the server. 

\section{Torque-Operator and Platform Description}\label{sec:architecture}
We firstly illustrate the design of our platform architecture, then describe the structure of Torque-Operator. Torque-Operator is written in Golang programming language. \textit{Singularity} \cite{Kurtzer2017SingularitySC} is the runtime container of our choice. Singularity is starting to be applied in many HPC centres \cite{Hu2019}, as it provides a secure means to capture and distribute software and  computer environment. For example, execution of a Singularity container only demands a user privilege, while a Docker container \cite{10.5555/2898929}, which is a container runtime widely adopted in cloud systems, requires root permission. Kubernetes supports Docker by default, though it can be adjusted to perform services for Singularity by adding Singularity-CRI \cite{Singularitycri}. Table~\ref{table:tool_list} manifests the list of core applications that construct the testbed.

\begin{table}[ht]
	\centering  % used for centering table	
	\resizebox{0.9\linewidth}{!}{%
	\begin{tabular}{| l| l|} % centered columns (3 columns)
		%\hline\hline                        %inserts double horizontal lines
		%\\ [0.5ex] % inserts table         
		\hline \hline%inserts single line
         Orchestrator & Kubernetes, Torque\\
		\hline
   	    Container runtime \& its support & Singularity, Singulairy-CRI\\
	 	\hline
		Operator & Torque-Operator \\
		\hline
		Compiler & Golang compiler\\[1ex] 		
		\hline\hline 
		%\\ [0.5ex] 
	\end{tabular}
}
	\caption[]{The list of core applications for the testbed.}\label{table:tool_list}
\end{table}

\subsection{Platform Architecture}
\begin{figure}[!t]
	\centering
	\includegraphics[width=0.50\textwidth]{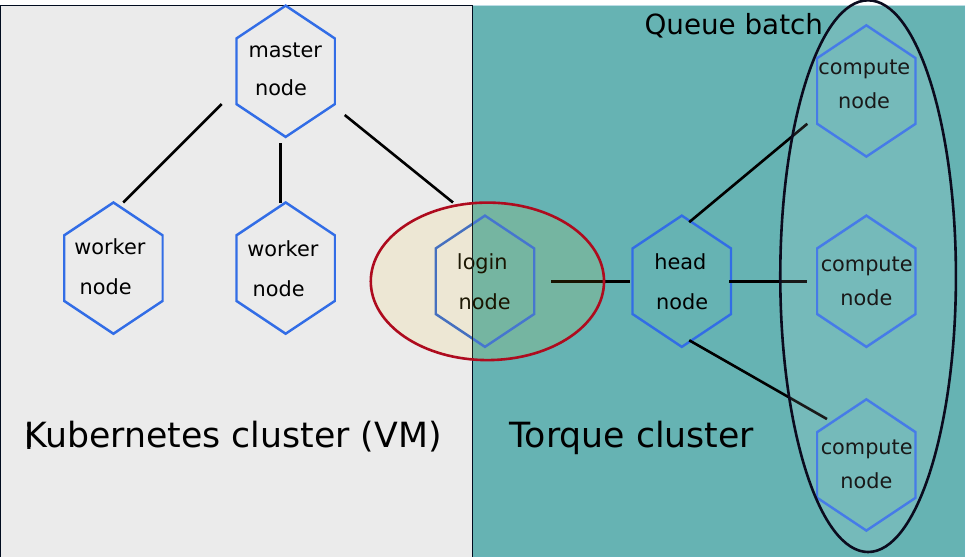}
	\caption[]{Architecture of the testbed. The login node belongs to both Kubernetes and Torque clusters. One queue (named batch) is shown in the Torque cluster.}
	\label{fig:arch_testbed}
\end{figure}

The architecture of our platform is designed to serve as the testbeds for the EU research project CYBELE\footnote{CYBELE:  Fostering Precision Agriculture and Livestock Farming through Secure Access to Large-Scale HPC-Enabled Virtual Industrial Experimentation Environment Empowering Scalable Big Data Analytics https://www.cybele-project.eu/}. The platform is composed of an HPC cluster with Torque as its workload manager and a big data cluster with Kubernetes as its orchestrator. Its architecture is illustrated in Fig.~\ref{fig:arch_testbed}. Noting that Fig.~\ref{fig:arch_testbed} is for illustration purpose, the number of nodes and the queues can vary in the testbeds.
 
\begin{figure}[!t]
	\centering
	\includegraphics[width=0.20\textwidth]{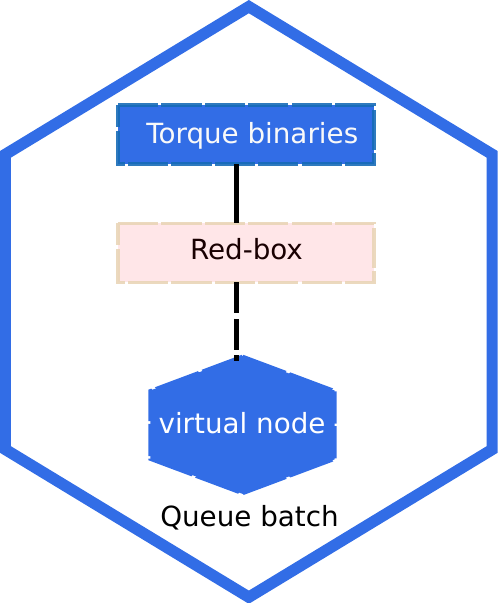}
	\caption[]{Architecture of Torque-Operator. This is the internal architecture of the login node as in Fig.~\ref{fig:arch_testbed}. The virtual node corresponds to the Torque queue (named batch) in Fig.~\ref{fig:arch_testbed}}
	\label{fig:arch_toruqe_operator}
\end{figure}

In Torque, nodes are grouped into queues. Each queue is associated with resources limits such as walltime, job size. One node can be included in multiple queues. The HPC cluster is composed of a head node which controls the whole cluster nodes and compute nodes which perform computation. The Torque login node in Fig.~\ref{fig:arch_testbed} also serves as one of the worker nodes in the Kubernetes cluster. The Kubernetes cluster incorporates a master node which schedules the jobs and worker nodes which execute the jobs. A virtual node indicated in Fig.~\ref{fig:arch_toruqe_operator} transfers Torque jobs to the Torque cluster. The Torque job submitted from the Kubernetes login node is scheduled by Kubernetes master node to the virtual node. The virtual node transfers the abstracted Torque jobs to the Torque queue through the Torque login node. The merits of this architecture are: 1) it provides users with flexibility to run containerised and non-containerised jobs, 2) the containerised applications can be better scheduled to Torque cluster by taking advantage of the scheduling policies of Kubernetes.

\subsection{Structure of Torque-Operator}
The Torque job script is encapsulated into a Kubernetes yaml job script. The yaml script is submitted from a Kubernetes login node (in our case, the login node is also the master node). The PBS script part is processed by Toque-Operator. A dummy pod is generated to transfer the Torque job specification to a scheduling queue (\textit{e.g.} waiting queue, test queue, which is a concept in the job scheduler). Torque-Operator invokes the Torque binary \textit{qsub} which submits PBS job to the Torque cluster. When the Torque job completes, Torque-operator creates a Kubernetes pod which redirects the results to the directory that the user specifies in the yaml file. 

As in WLM-Operator (Section~\ref{sec:related_work}), Torque-Operator includes a service program red-box. Red-box generates a Unix socket which allows data exchange among the Kubernetes and Torque processes. Torque-Operator introduces a new \textit{object kind} \textit{i.e. Torquejob} (\textit{Slurmjob} in WLM-Operator) and sets it as \textit{Kubernetes deployment}. Torque-Operator builds four Singularity containers which are deployed by Kubernetes on its worker nodes to perform the corresponding services, \textit{e.g.} create dummy pod to transfer the results from Torque to Kubernetes.

 \section{Test Case}\label{sec:resulst}
Simple experiments have been conducted to validate Torque-Operator. Fig.~\ref{fig:yaml_script} presents a Kubernetes yaml job script ($cow\_job.yaml$). More specifically, inside the yaml script, the Torque script requests 30-minute walltime and one compute node. The error file and output file are stored in $low.err$ and $low.out$ which locate in {\footnotesize $\$HOME/$} directory. The script appends the path $/usr/local/bin$ where Singularity binary resides. The Singularity container image $lolcow\_latest.sif$ is executed. The results are given in Fig.~\ref{fig:result_cow_job}. The user can view the status of the job easily from Kubernetes login node as shown in Fig.~\ref{fig:job_status}. Additionally, the status of the PBS job can be output using the Torque commands on the Torque login node.

 \begin{figure}
 	\begin{lstlisting}
 apiVersion: wlm.sylabs.io/v1alpha1
 kind: TorqueJob
 metadata
   name: cow
 spec:
   batch: |
   #!/bin/sh      
   #PBS -l walltime=00:30:00   
   #PBS -l nodes=1
   #PBS -e $HOME/low.err
   #PBS -o $HOME/low.out  
   export PATH=$PATH:/usr/local/bin 
   singularity run lolcow_latest.sif
 results:
   from: $HOME/low.out
   mount:
    name: data
    hostPath:
      path: $HOME/
      type: DirectoryOrCreate
 	\end{lstlisting} 
 	\vspace{-6pt}	
 	{\footnotesize
 	\begin{verbatim} 	
 	 $kubectl apply -f $HOME/cow_job.yaml 	 
 	\end{verbatim}
 }

 	\caption{An example of the yaml script and its submission command. The scirpt encloses a PBS script.}\label{fig:yaml_script}
 \end{figure}
%\vspace{-6pt}

 \begin{figure}
 	{\footnotesize
 		\begin{verbatim}	   
 		$kubectl get torquejob
 		NAME   AGE   STATUS
 		cow    2s   running 		
 		\end{verbatim}
 		%\vspace{-10pt}
 		\caption{The command to view the status of the yaml job}\label{fig:job_status}
 	}
 
 \end{figure}

\begin{figure}
	
	{\footnotesize
	\begin{verbatim}
 If one cannot enjoy reading a book over \
| and over again, there is no use in      |
| reading it at all.                      |
|                                         |
\ -- Oscar Wilde                          /
-----------------------------------------
\   ^__^
 \  (oo)\_______
    (__)\       )\/\
        ||----w |
        ||     ||

	\end{verbatim}
	
	\caption{A result of the Singularity job}\label{fig:result_cow_job}
}
\end{figure}

\section{Conclusion and Future Work}\label{sec:future_work}
We described the testbed architecture for the EU research project CYBELE and introduced the structure of Torque-Operator that extends WLM-Operator with Torque support. This testbed architecture creates a connection between HPC and cloud clusters. Moreover, it provides users with flexibility to run containerised and non-containerised jobs and may enhance the capability of container scheduling on HPC.    

The future work will focus on optimization of Torque-Operator that can offer more stable deployments. Performance evaluation will be carried out to compare efficiency of scheduling the container jobs by Kubernetes and Torque. The pilots of CYBELE project will be adopted as the benchmarks.

\section*{Acknowledgment}
This project has received funding from the European Union's Horizon 2020 research and innovation programme under grant agreement NO.825355.

% or


%\appendices
%% you can choose not to have a title for an appendix
%% if you want by leaving the argument blank
%\section{}
%% use section* for acknowledgement
%\section*{Acknowledgment}

\ifCLASSOPTIONcaptionsoff
  \newpage
\fi

\end{document}